# Machine learning based identification of buried objects using sparse whitened NMF

Ali Gharamohammadi
Electrical engineering
Sharif University of technology
Tehran, Iran
Aligharamohammadi@gmail.com

Fereidoon Behnia
Electrical engineering
Sharif University of technology
Tehran, Iran
behnia@sharif.edu

Arash Shokouhmand
Lane Department of Computer Science
and Electrical Engineering
West Virginia University
Morgantown, USA
as0357@mix.wvu.edu

*Abstract*—In this paper, a whitening-based algorithm has been applied to sparse non-negative matrix factorization (NMF) as a preprocessing practice enhancing the identification of buried object, significantly. In fact, without utilizing a suitable whitening algorithm, the input signals are very similar to one another. Therefore, conventional approaches are not able to identify and properly discriminate desired buried objects. A suitable whitening algorithm makes the covariance matrix diagonal, such that the processed signals highly resembles the unprocessed versions. Such a similarity would be helpful to identify new received signal. In this study, we have employed the zero-phase component analysis (ZCA) as our whitening method. Hence, although signals of different targets are very similar in many cases, the mentioned advantages, achieved by whitening method, make identification of targets much easier.

## I. Introduction

Buried objects identification (BOI) scheme is used to detect targets such as pipelines, landmines, cables, treasure and mines[1-8]. Detection of buried objects with a low depth has been a hot topic in recent research. In landmine detection problem, detection of all buried objects is of crucial importance. Hence, a method being able to detect all buried objects, even with some false alarms, should be a suitable one.

To identify shallow buried objects, ground surface reflection removal is very important . In the BOI problem, the reflected signal from the ground surface is usually much stronger than the one reflected from the buried objects, making shallow BOI difficult [4]. In this paper, the time gating method presented in [9] is utilized to circumvent the mentioned difficulty to some extent. This method itself relies on ultrashort/ultra wideband signals transmitted to and received from the medium under consideration.

Ultra-wideband (UWB) signaling has been widely utilized in recent investigations for more or less the same reason in diverse applications [10-11]. For example, finding tumors in breast cancer is a famous problem in which the researchers have widely utilized UWB concept [10].

Sparse NMF algorithm is well known in speech signal identification and widely adopted in the literature [12-13]. The problem of finding buried objects when the environment is sparse is like finding the source of a speech signal.

The major goal of this paper is to detect buried objects based on prior observations. In this paper, sparse NMF algorithm, used as an identification algorithm is modified by ZCA whitening [14]. Combination of whitening and sparse NMF is a novel idea which is proposed in this paper.

The remaining parts of this paper are organized as follows. In section II. A. Sparsity and non-negative matrix factorization are described. Besides, the whitening algorithm utilized in the paper is discussed in this section, which is of great significance in this work. In section III, experimental data acquisition and database are explained in detail. In section IV, the results of different methods are compared. Section V concludes the paper.

## II. Methodology

### A. Sparsity and non-negative matrix factorization

One of the popular methods to analyze non-negative data is non-negative matrix factorization (NMF) which can be used in a wide range of applications. The main goal of using this technique is to factor a non-negative data matrix to the product of two non-negative matrices. In the following, we show the main and the resultant matrices as:

$$M \in R^{F \times T}$$
$$W \in R^{F \times M} \quad (1)$$
$$H \in R^{M \times T}$$

In audio signal processing, this method is usually applied to the magnitude or the power spectrogram of the received signals. In this application, W and H are referred to as dictionary, and the activation matrix, respectively.

Whereas Both W and H are optimized in the unsupervised application, in the supervised application, W is fixed. In unsupervised application, the cost function is described as follows.

$$M \in R^{F \times T}$$
$$\overline{W}, \overline{H} = \underset{W,H}{\mathrm{argmin}}\, D(M|WH) + \mu |H| \quad (2)$$

where D is the cost function to be minimized.. In this paper, the β-divergence, $D_\beta$, is defined as follow.

$$D_\beta(x|y) \triangleq \begin{cases} \frac{1}{\beta(\beta-1)}(x^\beta - y^\beta - \beta y^{\beta-1}(x-y)) & \text{if } \beta \in R\setminus\{0,1\} \\ x(\log x - \log y) + (y-x) & \text{if } \beta = 1 \\ \frac{x}{y} - \log\frac{x}{y} - 1 & \text{if } \beta = 0 \end{cases} \quad (3)$$

Where, β=0 yields Itakura-Saito (IS) distance, β=1 yields the generalized Kullback-Leibler (KL) divergence, and β=2 yields the Euclidean distance. In this paper, KL divergence is used as the cost function. In supervised applications, due to the fact that W is fixed, the problem is defined as follow.

$$H = \underset{H}{\arg\min}\, D(M|WH) + \lambda|H| \quad (4)$$

In this work, W is considered as a fixed matrix. In [12] to update H, a simple algorithm is defined as the follows.

$$H \longleftarrow H \otimes \frac{W^T(M \otimes \Lambda^{\beta-2})}{W^T \Lambda^{\beta-1}} \quad (5)$$

where $\Lambda := WH$. For more information, refer to [12-13].

### B. Whitening

A kind of preprocessing method leading to orthogonality is whitening. In order to find an optimum solution, different criteria can be considered. In this paper, the covariance matrix and the correlation between signals before and after processing are used to compare different algorithms. Firstly, the covariance matrix should be almost diagonal, and along with that, every signal before and after processing should be correlated. To this end, there are several methods, some of which are introduced in [14], and a comparison has been made for their performance. For the input signal, x, mean and variance are represented as follows.

$$\begin{aligned} x &= (x_1,...,x_d)^T \\ E(x) &= \mu = (\mu_1,...,\mu_d)^T \\ \text{var}(x) &= \Sigma \end{aligned} \quad (6)$$

In these methods, eigendecomposition and polar decomposition are stated as (7).

$$\begin{aligned} \Sigma &= u\Lambda u^T \\ \Sigma &= V^{1/2}\Lambda V^{1/2} \\ P &= G\Theta G^T \end{aligned} \quad (7)$$

The following table describes different methods of formulations.

TABLE I. FORMULATION OF DIFFERENT METHODS IN WHITENING ALGORITHM

| Method | Whitening matrix |
|---|---|
| PCA | $\Sigma^{-1/2}$ |
| ZCA | $\Lambda^{-1/2}u^T$ |
| ZCA-cor | $P^{-1/2}V^{-1/2}$ |
| PCA-cor | $\Theta^{-1/2}G^T V^{-1/2}$ |

## III. DATA ACQUISITION

Experimental data has been collected in a lab at Georgia Tech University. In what follows, the experimental setup is described. Also, for more information and details see [1].

The system consists of an antenna array consisting of similar antennas. A vector network analyzer and 3-D positioner are also used to scan the region. The region size is 120*120 cm with fixed height and 2 cm scan step in x and y-axes. This setup is depicted in Fig. 1. The frequency bandwidth is approximately 8 GHz scanned in 20 MHz steps. Meanwhile, permittivity of soil is 4, and In this paper, only T1R1 pair is considered for sake of simplicity. The scenario used in this paper is depicted in Fig. 2.

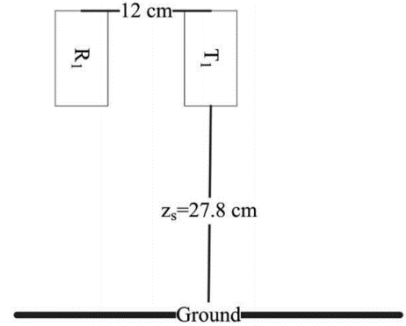

Fig. 1. The first transmitter and first receiver in the scenario as used in this paper [1].

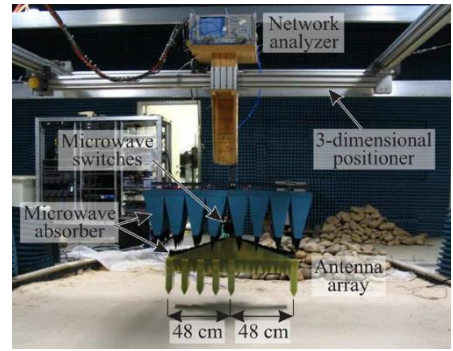

Fig. 2. Experimental setup [1].

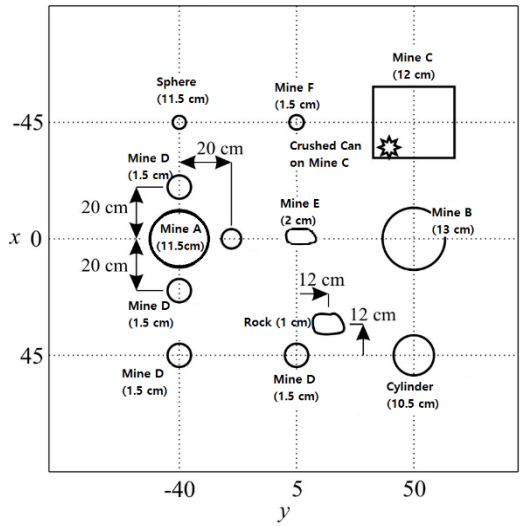

Fig. 3. The implemented scenario including name, location, and depth of targets in the x-y scene.

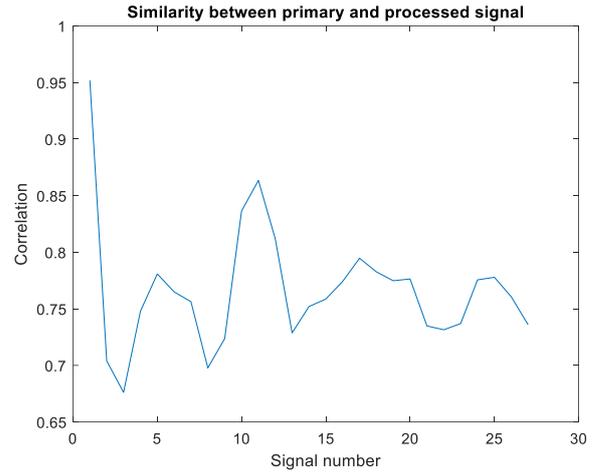

(c)

Fig. 4. ZCA whitening algorithm results. (a) the covariance matrix of input signals. (b) the covariance matrix of processed signals by ZCA. (c) correlation between the first input signal and the processed signals. A number is assigned to each signal in accordance with table 2. It is noticeable that the 12[th] signal is related to vacant(no target) scenario.

## IV. RESULTS AND DISCUSSION

In this paper, several whitening algorithms as presented in [14] are implemented. There are two main notes which should be considered in whitening results. Firstly, the covariance matrix should be diagonal. Secondly, the processed signal should be very similar to the their input versions. ZCA indicates better performance compared to other methods in terms of the mentioned points [14]. Therefore, ZCA has been selected as the whitening algorithm. As seen in Fig. 4, not only the covariance matrix has been improved significantly from the viewpoint of diagonality, but also the input and the processed signals are very similar according to correlation results.

By applying whitening algorithm we expect the results of the Sparse NMF algorithm to be improved. In the test scenario, there is only one repetitive object which is mine D. For this reason, mine D can be used in order to test the algorithm. Applying the whitening algorithm, mine D is detected as a buried object when the scanner searches the region around. Without whitening algorithm, Sparse NMF is not able to find the buried objects from the dictionary truly.

## V. CONCLUSION

In this paper, the sparse NMF algorithm has been modified by the whitening algorithm. The results of whitening algorithm demonstrate that the data after processing are uncorrelated with the unprocessed input signal. Besides, the covariance matrix is diagonal. The processed data by ZCA whitening is truly converged in sparse NMF algorithm. Also, using whitening with other methods such as correlation can be beneficial. This methodology, especially for sparse environment, has significant results.

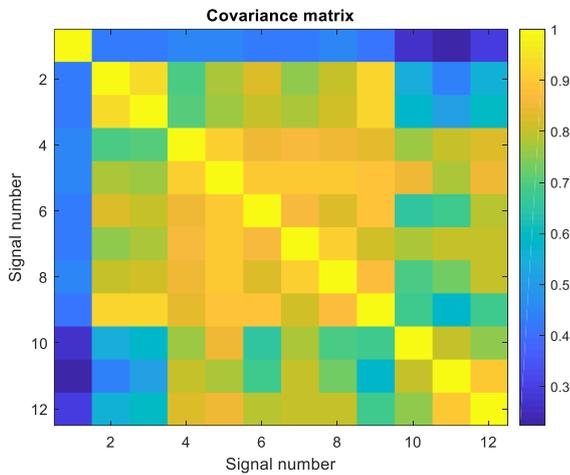

(a)

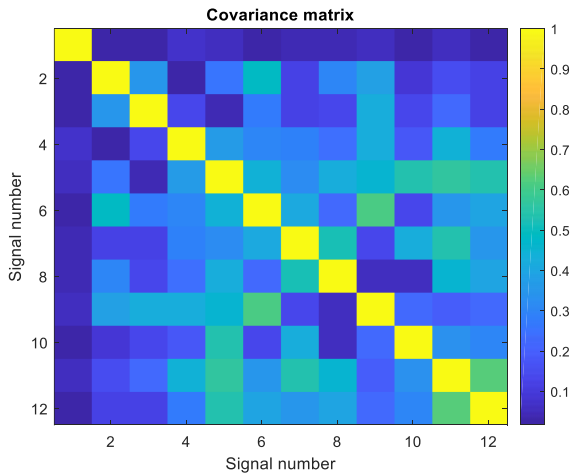

(b)

TABLE II.  PROPERTIES OF BURIED OBJECTS IN EXPERIMENTAL SETUP[1]. AT AND AP ARE AN ABBREVIATION OF ANTI-TANK AND ANTI-PERSONNEL RESPECTIVELY.

| Target | properties | | | |
|---|---|---|---|---|
| | Type | Material | Dimensions (cm) | Number of signal |
| Mine A | AT | Plastic | 22.2(D), 9.2(H) | 5 |
| Mine B | AT | Plastic | 24(D), 12(H) | 7 |
| Mine C | AT | Plastic | 31.2(L), 27.5(W), 11.3(H) | 3 |
| Mine D | AP | Plastic | 9(D), 4.5(H) | 10 |
| Mine E | AP | Plastic | 11.9(L), 6.4(W), 2(H) | 6 |
| Mine F | AP | Plastic | 5.6(D), 4(H) | 2 |
| Mine simulant | AP | Plastic | 7.5(D), 3.8(H) | 8 |
| Sphere | Buried clutter | Aluminum | 5.1(D) | 1 |
| Rock | Buried clutter | Rock | 12(L), 8(W), 7.5(H) | 9 |
| Crushed Can | Buried clutter | Aluminum | 8(D), 3(H) | 4 |
| Cylinder | Buried clutter | Nylon | 15.5(D), 7.6(H) | 11 |

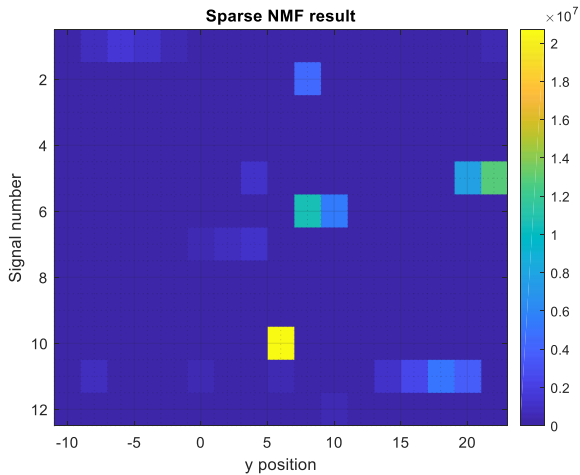

(a)

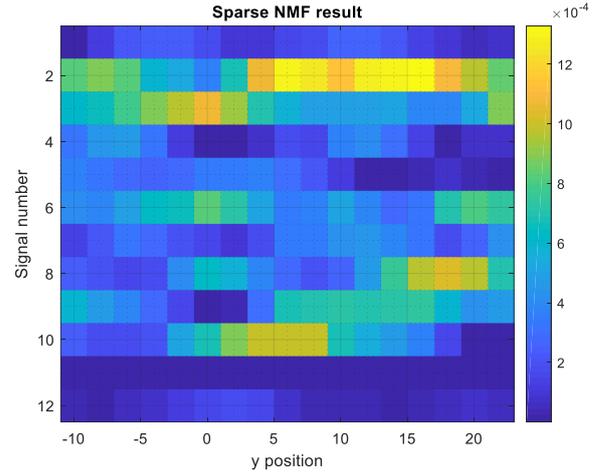

(b)

Fig. 5. The effect of applying the whitening algorithm signal of mine D(10). (a) without whitening. (b) with whitening.